\RequirePackage{ifpdf}
\ifpdf 
\documentclass[pdftex]{sigma}
\else
\documentclass{sigma}
\fi

\usepackage{mathrsfs}

\def\bbbr{{\Bbb R}}

\def\bbbz{{\Bbb Z}}
\def\openone{\leavevmode\hbox{\small1\kern-3.3pt\normalsize1}}

\def\openone{\leavevmode\hbox{\small1\kern-3.3pt\normalsize1}}

\def\diag{\mbox{diag\,}}

\def\bbbr{{\Bbb R}}

\def\bbbz{{\Bbb Z}}

\begin{document}

\numberwithin{equation}{section}

\allowdisplaybreaks

\renewcommand{\PaperNumber}{039}

\FirstPageHeading

\renewcommand{\thefootnote}{$\star$}

\ShortArticleName{Soliton Solutions for $N$-wave Equations with
$\mathbb{Z}_2$ and $\mathbb{Z}_2\times\mathbb{Z}_2$ Reductions}

\ArticleName{$\boldsymbol{N}$-Wave Equations with Orthogonal Algebras:\\
$\boldsymbol{\mathbb{Z}_2}$ and
$\boldsymbol{\mathbb{Z}_2\times\mathbb{Z}_2}$ Reductions and
Soliton Solutions\footnote{This paper is a contribution to the
Vadim Kuznetsov Memorial Issue ``Integrable Systems and Related
Topics''. The full collection is available at
\href{http://www.emis.de/journals/SIGMA/kuznetsov.html}{http://www.emis.de/journals/SIGMA/kuznetsov.html}}}

\Author{Vladimir S. GERDJIKOV~$^\dag$, Nikolay A. KOSTOV~$^{\dag
\ddag}$ and Tihomir I. VALCHEV~$^\dag$}

\AuthorNameForHeading{V.S. Gerdjikov, N.A. Kostov and T.I. Valchev}

\Address{$^\dag$~Institute for Nuclear Research and Nuclear
Energy, Bulgarian Academy of Sciences,\\
$\phantom{^\dag}$~72 Tsarigradsko chaussee, 1784 Sofia, Bulgaria}
\EmailD{\href{mailto:gerjikov@inrne.bas.bg}{gerjikov@inrne.bas.bg}}

\Address{$^\ddag$~Institute of Electronics, Bulgarian Academy of
Sciences,\\
$\phantom{^\ddag}$~72 Tsarigradsko chaussee, 1784 Sofia, Bulgaria}

\ArticleDates{Received November 21, 2006, in f\/inal form February
08, 2007; Published online March 03, 2007}

\Abstract{We consider $N$-wave type equations related to the
orthogonal algebras obtained from the generic ones via additional
reductions. The f\/irst $\mathbb{Z}_2$-reduction is the canonical
one. We impose a second $\bbbz_2$-reduction and consider also the
combined action of both reductions. For all three types of
$N$-wave equations we construct the soliton solutions by
appropriately modifying the Zakharov--Shabat dressing method. We
also brief\/ly discuss the dif\/ferent types of one-soliton
solutions. Especially rich are the types of one-soliton solutions
in the case when both reductions are applied. This is due to the
fact that we have two dif\/ferent conf\/igurations of eigenvalues
for the Lax operator $L$: doublets,  which consist of pairs of
purely imaginary eigenvalues, and quadruplets. Such situation is
analogous to the one encountered in the sine-Gordon case, which
allows two types of solitons: kinks and breathers. A new physical
system, describing Stokes-anti Stokes Raman scattering is
obtained. It is represented by a~$4$-wave equation related to the
${\bf B}_2$ algebra with a canonical $\mathbb{Z}_2$ reduction.}


\Classification{37K15; 17B70; 37K10; 17B80}

\section{Introduction}\label{sec:1}

The $N$-wave equation related to a semisimple Lie algebra
$\mathfrak{g}$ is a matrix system of nonlinear dif\/ferential
equations of the type
\begin{gather}\label{$N$-waves}
i[J,Q_{t}(x,t)]-i[I,Q_{x}(x,t)]+[[I,Q(x,t)],[J,Q(x,t)]]=0,
\end{gather}
where the  squared brackets denote the commutator of matrices and
the subscript means a partial derivative with respect to
independent variables $x$ and $t$. The constant matrices $I$ and
$J$ are regular and real elements of the Cartan subalgebra
$\mathfrak{h}$ of the Lie algebra $\mathfrak{g}$. The
matrix-valued function $Q(x,t)\in\mathfrak{g}$ can be expanded as
follows
\[Q(x,t)=\sum_{\alpha\in\Delta}Q_{\alpha}(x,t)E_{\alpha},\]
where $\Delta$ denotes the root system and $E_{\alpha}$ are
elements of the Weyl basis of $\mathfrak{g}$ parametrized by roots
of $\mathfrak{g}$. It is also assumed that $Q(x,t)$ satisf\/ies a
vanishing boundary condition, i.e.\
$\lim\limits_{x\to\pm\infty}Q(x,t)=0$.

The $N$-wave equation is an example of an S-integrable
evolutionary equation. Such type of equations appear in nonlinear
optics and describes the propagation of $N$ wave packets in
nonlinear media \cite{bloem}.

Another application of the $N$-wave systems is in dif\/ferential
geometry. Ferapontov \cite{fer} showed that $N$-wave equations
naturally occurred when one studied isoparametric hypersurfaces in
spheres.

Our aim in this paper is two-fold. First, we outline the
derivation of $N$-wave equations with $\bbbz_2$-reductions and
calculate their soliton solutions by applying one of the basic
methods in theory of integrable systems -- Zakharov--Shabat
dressing procedure. One of the considered examples has a physical
interpretation -- it describes a Stokes-anti-Stokes Raman
scattering in nonlinear optics. Secondly, we further reduce the
$N$-wave equations by imposing a second $\bbbz_2$-reduction. As a
result we derive a special class of $N$-wave equations which, like
the sine-Gordon equation~(SG), possess breather type solutions.
Next we obtain their soliton solutions. The additional symmetries
of the nonlinear equations have been taken into account when we
chose a proper dressing factor. The soliton solutions of a
$\mathbb{Z}_2\times\mathbb{Z}_2$-reduced $N$-wave equation are of
two types. The solitons of  the f\/irst type are connected with a
couple of purely imaginary discrete eigenvalues of the Lax
operator $L$. These solutions correspond to the kinks (or
topological solitons) of~SG. The solitons of the second  type,
associated with quadruplets of discrete eigenvalues $L$ which are
symmetrically located with respect to the coordinate axis in the
complex $\lambda$-plane, are analogues of the breather solutions
of~SG. Recently the Darboux-dressing transformations have been
applied to the Lax pair associated with systems of coupled
nonlinear wave equations: vector nonlinear Schr\"odinger-type
equations \cite{dl07}. Solutions with boomeronic and trapponic
behaviour were investigated.

The problem for classif\/ication and investigation of all
admissible reductions of an integrable equation is one of the
fundamental problems in the theory of integrable systems. $N$-wave
equations with canonical $\mathbb{Z}_2$ symmetries have been
discussed for the f\/irst time by Zakharov and Manakov in
\cite{zak-man} for $\mathfrak{g}\simeq\mathfrak{sl}(n)$. More
recently the $\bbbz_2$-reductions of the $N$-wave equations
related to the low-rank simple Lie algebras were analyzed  and
classif\/ied~\cite{GGIK,GGK}.

Since we shall deal with the inverse scattering  transform we
begin with a reminder of all necessary facts concerning that
theory. For more detailed information we recommend the
monographs~\cite{blue-bible} and~\cite{brown-bible}.

\section{General formalism}\label{sec:2}

As we mentioned above the $N$-wave system (\ref{$N$-waves}) is an
integrable one. It admits a Lax representation with Lax operators
\begin{gather}
L\psi(x,t,\lambda)= \left(i\partial_x+U(x,t,\lambda)\right)\psi(x,t,\lambda)=0 ,\nonumber\\
M\psi(x,t,\lambda)=\left(i\partial_t+V(x,t,\lambda)\right)\psi(x,t,\lambda),\label{Lax1}
\end{gather}
where $\lambda$ is an auxiliary (so-called spectral) parameter and
the potentials $U(x,t,\lambda)$, $V(x,t,\lambda)$ are elements of
$\mathfrak{g}$ and they are linear functions of $\lambda$
def\/ined by
\begin{gather*}
U(x,t,\lambda)=U^0(x,t)-\lambda J=[J,Q(x,t)]-\lambda J, \\
V(x,t,\lambda)=V^0(x,t)-\lambda I=[I,Q(x,t)]-\lambda I.
\end{gather*}
The nonlinear evolution equation itself is equivalent to the
compatibility condition of the dif\/fe\-ren\-tial operators $L$
and $M$
\[
[L,M]=0\quad\Leftrightarrow\quad
i[J,Q(x,t)_{t}]-i[I,Q(x,t)_{x}]+[[I,Q(x,t)],[J,Q(x,t)]]=0.
\]
Since $L$ and $M$ commute they have the same eigenfunctions and
the system (\ref{Lax1}) can be presented~by
\begin{gather}
L\psi(x,t,\lambda)= \left(i\partial_x+U(x,t,\lambda)\right)
\psi(x,t,\lambda)=0,\nonumber \\
M\psi(x,t,\lambda)=\left(i\partial_t+V(x,t,\lambda)\right)
\psi(x,t,\lambda)=\psi(x,t,\lambda)C(\lambda),\label{Lax2}
\end{gather}
where $C(\lambda)$ is a constant matrix with respect to $x$ and
$t$. The fundamental solutions $\psi(x,t,\lambda)$ of the
auxiliary linear system (\ref{Lax2}) take values in the Lie group
$G$ which corresponds to the Lie algebra $\mathfrak{g}$.

In order to f\/ind the soliton solutions we need the so-called
fundamental analytic solutions (FAS) derived for the f\/irst time
in \cite{brown-bible} for $G\simeq SL(n)$; for other simple Lie
algebras see \cite{IP2,gerdji}. There is a standard algorithm to
construct these solutions by using another class of fundamental
solutions of the linear problem (\ref{Lax2}) -- Jost solutions.
The Jost solutions $\psi_{\pm}(x,t,\lambda)$ are determined by
their asymptotics at inf\/inity, i.e.
\[
\lim_{x\to\pm\infty}\psi_{\pm}(x,t,\lambda)e^{i\lambda
Jx}=\openone.
\]

\noindent {\bf Remark.} This def\/inition is correct provided we
f\/ixed up the matrix-valued function $C(\lambda)$ by
\[
C(\lambda)=\lim_{x\to\pm\infty}V(x,t,\lambda)=-\lambda I,
\]
i.e.\ the asymptotics of $\psi_{\pm}(x,t,\lambda)$ are
$t$-independent. $C(\lambda)$ is directly related to the
dispersion law of the nonlinear equation. Thus the dispersion law
of the $N$-wave equation is a linear function of the spectral
parameter $\lambda$.

\medskip

The Jost solutions $\psi_{\pm}(x,t,\lambda)$ are linearly related,
which means that there exists a matrix $T(t,\lambda)$ such that
\[
\psi_{-}(x,t,\lambda)=\psi_{+}(x,t,\lambda)T(t,\lambda).
\] The
matrix-valued function $T(t,\lambda)$ is called a scattering
matrix. Its time evolution is determined by the second equation of
(\ref{Lax2}), i.e.
\[
i\partial_t T(t,\lambda)-\lambda[I,T(t,\lambda)]=0.
\]
Consequently
\begin{gather}\label{T_evol}
T(t,\lambda)=e^{-i\lambda It}T(0,\lambda)e^{i\lambda It}.
\end{gather}

The inverse scattering transform (IST) allows one to solve the
Cauchy  problem for the non\-li\-near evolution equation, i.e.\
f\/inding a solution $Q(x,t)$ when its initial condition $Q_{\rm
in}(x)=Q(x,0)$ is given. The idea of IST is illustrated in the
following diagram
\[
Q_{\rm in}(x)\to U(x,0,\lambda)\stackrel{\rm DSP} \longrightarrow
T(0,\lambda) \longrightarrow T(t,\lambda)\stackrel{\rm
ISP}\longrightarrow U(x,t,\lambda)\to Q(x,t).
\] The f\/irst step consists in constructing the scattering
matrix at some initial moment $t=0$ by using the potential at the
same moment (or equivalently by the solution $Q_{\rm in}(x)$ at
that moment). This is a direct scattering problem (DSP). The
evolution of the scattering matrix (data) is already known and it
is given by (\ref{T_evol}). The third step is recovering of the
potential $U(x,t,\lambda)$ and respectively the solution of the
nonlinear equation $Q(x,t)$ at an arbitrary moment from the
scattering data at that moment~-- this is an inverse scattering
problem (ISM). That step is actually the only nontrivial one. Thus
following all steps we can solve the Cauchy problem for the
nonlinear evolution equation. Since we know the evolution of
scattering data we can easily determine the time dependence of
fundamental solutions, solutions of nonlinear problem etc.

The Jost solutions are def\/ined for real values of $\lambda$ only
(they do not possess analytic properties for
$\lambda\notin\mathbb{R}$). For our purpose it is necessary to
construct fundamental solutions which admit analytic continuation
beyond the real axes. It can be shown that there exist fundamental
solutions $\chi^{+}(x,t,\lambda)$ and $\chi^{-}(x,t,\lambda)$
analytic in the upper half-plane $\mathbb{C}_{+}$ and in the lower
half-plane $\mathbb{C}_{-}$ of the spectral parameter
respectively. They can be obtained from the Jost solutions by a
simple algebraic procedure proposed by Shabat \cite{Sha,Sha1}, see
also \cite{brown-bible}. The procedure uses a Gauss decomposition
of the scattering matrix $T(t,\lambda)$, namely
\[
\chi^{\pm}(x,t,\lambda)=\psi_{-}(x,t,\lambda)S^{\pm}(t,\lambda)=
\psi_{+}(x,t,\lambda)T^{\mp}(t,\lambda)D^{\pm}(\lambda),
\] where
matrices $S^{\pm}(t,\lambda)$, $T^{\pm}(t,\lambda)$ and
$D^{\pm}(\lambda)$ are Gauss factors of the matrix $T(t,\lambda)$,
i.e.
\[
T(t,\lambda)=T^{\mp}(t,\lambda)D^{\pm}(\lambda)(S^{\pm}(t,\lambda))^{-1}.
\]
The matrices $S^{+}(t,\lambda)$ and $T^{+}(t,\lambda)$
(resp.~$S^{-}(t,\lambda)$ and $T^{-}(t,\lambda)$) are upper
(resp.~lower) triangular with unit diagonal elements and taking
values in $G$. Their time dependence is given~by:
\begin{gather*}
i\partial_t S^\pm(t,\lambda)-\lambda[I,S^\pm(t,\lambda)]=0, \qquad
i\partial_t T^\pm(t,\lambda)-\lambda[I,T^\pm(t,\lambda)]=0.
\end{gather*}
The matrices $D^{+}(\lambda)$ and $D^{-}(\lambda)$ are diagonal
and allow analytic extension in $\lambda$ for $\mbox{Im}\lambda
>0$ and $\mbox{Im}\lambda <0$. They do not depend on time and
provide the generating functionals of the integrals of motion of
the nonlinear evolution equation \cite{brown-bible,IP2,GGIK,GGK},
see also the review paper \cite{gerdji}.

A powerful method for obtaining solutions to nonlinear
dif\/ferential equations is B\"{a}cklund transformation (see
\cite{Mat} and \cite{Rog,rs02} for more detailed information). A
B\"acklund transformation maps a solution of a dif\/ferential
equation into a solution of another dif\/ferential equation. If
both equations coincide one speaks of an auto-B\"{a}cklund
transformation. A very important particular case of an
auto-B\"{a}cklund transformation is the dressing method proposed
by Zakharov and Shabat \cite{Zak-Shab}. Its basic idea consists in
constructing a new solution $Q(x,t)$ starting from a~known
solution $Q_0(x,t)$ taking into account the existence of the
auxiliary linear system~(\ref{Lax2}).

Let $\psi_0(x,t,\lambda)$ satisfy the linear problem
\begin{gather}\label{bare}
L_0\psi_0(x,t,\lambda)=i\partial_{x}\psi_0(x,t,\lambda)+([J,Q_0(x,t)]-\lambda
J) \psi_0(x,t,\lambda)=0.
\end{gather}
We construct a function $\psi(x,t,\lambda)$ by introducing a gauge
transformation $g(x,t,\lambda)$ -- dressing procedure of the
solution $\psi_0(x,t,\lambda)$
\[
\psi_0(x,t,\lambda)\to\psi(x,t,\lambda)=
g(x,t,\lambda)\psi_0(x,t,\lambda)
\] such that the linear system
(\ref{bare}) is covariant under the action of that gauge
transformation. Thus the dressing factor has to satisfy
\begin{gather}\label{factor PDE}
i\partial_{x}g(x,t,\lambda)+[J,Q(x,t)]g(x,t,\lambda) -
g(x,t,\lambda) [J,Q_0(x,t)]-\lambda[J,g(x,t,\lambda)]=0.
\end{gather}
If we choose a dressing factor which is a meromorphic function of
the spectral parameter $\lambda$ as follows
\begin{gather}\label{g}
g(x,t,\lambda)=\openone+\frac{A(x,t)}{\lambda-\lambda^{+}}
+\frac{B(x,t)}{\lambda-\lambda^{-}},
\end{gather}
where $\lambda^{\pm}\in\mathbb{C}_{\pm}$, we obtain the following
relation between $Q(x,t)$ and $Q_0(x,t)$
\[
[J,Q(x,t)]=[J,Q_0(x,t)+A(x,t)+B(x,t)].
\]
As a result we are able to construct new solutions if we know the
functions $A(x,t)$ and $B(x,t)$. We will show later how this can
be done.

The simplest class of solutions are the so-called ref\/lectionless
potentials. A soliton solution is obtained by dressing the trivial
solution $Q_0(x,t)\equiv 0$. Then the fundamental solution of the
linear problem is just a plane wave
$\psi_0(x,t,\lambda)=e^{-i\lambda(Jx+It)}$ and the one-soliton
solution itself is given by
\[
[J,Q_{1s}(x,t)]=[J,A_{1s}(x,t)+B_{1s}(x,t)].
\]

As we said above the dressing procedure maps a solution of the
linear problem to another solution of a linear problem with a
dif\/ferent potential. In particular the Jost solutions
$\psi_{0,\pm}(x,t,\lambda)$ are transformed into
\[
\psi_{\pm}(x,t,\lambda)=g(x,t,\lambda)\psi_{0,\pm}(x,t,\lambda)g^{-1}_{\pm}(\lambda),
\]
where the factors $g_{\pm}(\lambda)$ ensure the proper asymptotics
of the dressed solutions and are def\/ined~by
\[
g_{\pm}(\lambda)=\lim_{x\to\pm\infty}g(x,t,\lambda).
\]
Hence the dressed scattering matrix $T(t,\lambda)$ reads
\[
T(t,\lambda)=g_{+}(\lambda)T_{0}(t,\lambda)g^{-1}_{-}(\lambda).
\]
It can be proven that the FAS $\chi^{\pm}_0(x,t,\lambda)$
transform into
\begin{gather}\label{FAS_dressed}
\chi^{\pm}(x,t,\lambda)=g(x,t,\lambda)\chi^{\pm}_{0}(x,t,\lambda)g^{-1}_{-}(\lambda).
\end{gather}

The spectral properties of $L$ are determined by the behaviour of
its resolvent operator. The resolvent $R(t,\lambda)$ is an
integral operator (see  \cite{LMP,gerdji} for more details) given
by
\[R(t,\lambda)f(x)=\int^{\infty}_{-\infty}\mathcal{R}(x,y,t,\lambda)f(y)dy,\]
where the kernel $\mathcal{R}(x,y,t,\lambda)$ must be a piece-wise
analytic function of $\lambda$ satisfying the equation
\[L\mathcal{R}(x,y,t,\lambda)=\delta(x-y)\openone.\]
The kernel $\mathcal{R}(x,y,t,\lambda)$ can be constructed by
using the fundamental analytical solutions as follows
\[
\mathcal{R}(x,y,t,\lambda)=\left\{\begin{array}{cc}
\mathcal{R}^{+}(x,y,t,\lambda),& \mbox{Im}\lambda>0,\\
\mathcal{R}^{-}(x,y,t,\lambda),&
\mbox{Im}\lambda<0,\end{array}\right.
\]
where
\begin{gather*}
\mathcal{R}^{\pm}(x,y,t,\lambda)=\pm
i\chi^{\pm}(x,t,\lambda)\Theta^{\pm}(x-y)(\chi^{\pm}(y,t,\lambda))^{-1},\\
\Theta^{\pm}(x-y)=\theta(\mp(x-y))\Pi-\theta(\pm(x-y))(1-\Pi),\\
\Pi=\sum^{a}_{p=1}E_{pp}\ ,\qquad
(E_{pq})_{rs}=\delta_{pr}\delta_{qs},
\end{gather*}
where $\theta$ is the standard Heaviside function and $a$ is the
number of positive eigenvalues of $J$. Due to the fact that we
have chosen $J$ to be a real matrix with
\[
J_1>J_2>\cdots >J_a >0> J_{a+1} > \cdots >J_n,
\]
the resolvent $R(t,\lambda)$ is a bounded integral operator for
$\mbox{Im}\lambda\neq 0$. For $\mbox{Im}\lambda = 0$
$R(t,\lambda)$ is an unbounded integral operator, which means that
the continuous  spectrum of $L$ f\/ills up the real axes
$\mathbb{R}$. Since the discrete part of the spectrum of $L$ is
determined by the poles of $R(t,\lambda)$ it coincides with the
poles and zeroes of $\chi^{\pm}(x,t,\lambda)$.

From (\ref{FAS_dressed}) and from the explicit form of
$\mathcal{R}(t,\lambda)$ it follows that the dressed kernel is
related to the bare one by
\begin{gather*}
\mathcal{R}^{\pm}(x,y,t,\lambda)=g(x,t,\lambda)\mathcal{R}^{\pm}_0(x,y,t,\lambda)
g^{-1}(y,t,\lambda).
\end{gather*}
If we assume that the ``bare'' operator $L_0$ has no discrete
eigenvalues then the poles of $g(x,t,\lambda)$ determine the
discrete eigenvalues of $L$
\[
L_0\stackrel{g}{\longrightarrow}L\quad\Leftrightarrow\quad\mbox{spec}
(L_0)\quad
\to\quad\mbox{spec}(L)=\mbox{spec}(L_0)\cup\left\{\lambda^{+},\lambda^{-}\right\}.
\]

Many classical integrable systems correspond to Lax operators with
potentials possessing additional symmetries. That is why it is of
particular interest to consider the case when certain symmetries
are imposed on the potential $U(x,t,\lambda)$ (resp.\ on the
solution $Q(x,t)$).

Let $G_{R}$ be a discrete group acting in $G$ by group
automorphisms and in $\mathbb{C}$ by conformal transformations
\[\kappa: \lambda\to \kappa(\lambda).\]
Therefore we have an induced action of $G_{R}$ in the space of
functions $f(x,t,\lambda)$ taking values in~$G$ as follows
\[
\mathcal{K}:f(x,t,\lambda)\to\tilde{f}(x,t,\lambda)=
\tilde{K}\left(f\left(x,t,\kappa^{-1}(\lambda)\right)\right),
\qquad \tilde{K}\in\mathrm{Aut}(G).
\] This action in turn induces
another action on the dif\/ferential operators $L$ and $M$
\begin{gather*}
\tilde{L}(\lambda)=\mathcal{K}L(\kappa^{-1}(\lambda))\mathcal{K}^{-1},\\
\tilde{M}(\lambda)=\mathcal{K}M(\kappa^{-1}(\lambda))\mathcal{K}^{-1}.
\end{gather*}
The group action is consistent with Lax representation which is
equivalent to invariance of Lax representation under the action of
$G_R$, i.e.\ if $[L,M]=0$ then also
\[
[\tilde{L},\tilde{M}]=0.
\]

The requirement of $G_R$-invariance of the set of fundamental
solutions $\{\psi(x,t,\lambda)\}$ leads to the following symmetry
condition
\begin{gather*}
\tilde{K}U(x,t,\kappa^{-1}(\lambda))\tilde{K}^{-1}=U(x,t,\lambda).
\end{gather*}
In other words the potential $U(x,t,\lambda)$, as well as $Q(x,t)$
are reduced. This fact motivates the name of the group $G_R$ -- a
reduction group, proposed by Mikhailov in \cite{mik}.

One can prove that the dressing factor $g(x,t,\lambda)$ ought to
be invariant under the action of the reduction group $G_R$
\begin{gather}\label{invariance}
\tilde{K}\left(g(x,t,\kappa^{-1}(\lambda))\right)=g(x,t,\lambda).
\end{gather}

\section[$N$-wave equations with a $\mathbb{Z}_2$ reduction]{$\boldsymbol{N}$-wave equations with a $\boldsymbol{\mathbb{Z}_2}$ reduction}

In this chapter we are going to demonstrate an algorithm to derive
soliton solutions for a $N$-wave equation related to the
orthogonal algebra (i.e.\
$\mathfrak{g}\equiv\mathfrak{so}(n,\mathbb{C})$ ) with a
$\mathbb{Z}_2$ reduction imposed on~it. We pay special attention
to the particular case of
$\mathbf{B}_2\simeq\mathfrak{so}(5,\mathbb{C})$ algebra: we
display the explicit form of the $4$-wave system and calculate by
components its one-soliton solutions. One of the $4$-wave systems
under consideration enjoys a physical application in nonlinear
optics. We are following the ideas presented by Zakharov and
Mikhailov in \cite{zak-mik}.

\subsection[Physical applications of 4-wave equation with a $\mathbb{Z}_2$
reduction]{Physical applications of 4-wave equation with a
$\boldsymbol{\mathbb{Z}_2}$ reduction}

From now on we shall deal with the orthogonal group $SO(n,\mathbb{C})$ and its Lie
algebra $\mathfrak{so}(n,\mathbb{C})$. That is why we remind the
reader several basic facts concerning this topic without proving
them, for more detailed information we recommend the book
\cite{GG}.

The orthogonal group is, by def\/inition, a matrix group which
consists of all isometries acting on $\mathbb{C}^n$, i.e.\
\[
SO(n,\mathbb{C})=\left\{C\in GL(n,\mathbb{C})|\, C^TSC=S\right\},
\]
where $S$ stands for the metric in $\mathbb{C}^n$. It determines a
scalar product by the formula
\[(u,v)=u^TSv,\qquad u,v\in\mathbb{C}^n .\]
$\mathfrak{so}(n,\mathbb{C})$ is the matrix Lie algebra of all
inf\/initesimal isometries on $\mathbb{C}^n$, i.e.
\[
\mathfrak{so}(n,\mathbb{C})=\left\{\mathfrak{c}\in
\mathfrak{gl}(n,\mathbb{C})|\,
\mathfrak{c}^TS+S\mathfrak{c}=0\right\}.
\]
The orthogonal algebra is a semisimple Lie algebra. We remind that
its maximal commutative subalgebra is called a Cartan subalgebra
$\mathfrak{h}$. It is more convenient to work with a basis in
$\mathbb{C}^n$ such that the matrix of $S$ in that basis reads
\[S=\sum^n_{k=1}(-1)^{k-1}(E_{k,2n+1-k}+E_{2n+1-k,k})+
(-1)^n E_{nn}
,\qquad\mbox{for}\quad\mathbf{B}_n\simeq\mathfrak{so}(2n+1,\mathbb{C})\]
and
\[
S=\sum^n_{k=1}(-1)^{k-1}(E_{k,2n+1-k}+E_{2n+1-k,k}),
\qquad\mbox{for}\quad\mathbf{D}_n\simeq\mathfrak{so}(2n,\mathbb{C}).
\]
This choice of a basis ensures that the corresponding Cartan
subalgebras consist of diagonal matrices. More precisely, Cartan
basis $\{H_k\}^n_{k=1}$, i.e.\ the basis which spans
$\mathfrak{h}$, has the form
\begin{gather*}
H_k=E_{kk}-E_{2n+2-k,2n+2-k},\qquad\mbox{for}\quad\mathbf{B}_n,\\
H_k=E_{kk}-E_{2n+1-k,2n+1-k}, \qquad\mbox{for}\quad\mathbf{D}_n.
\end{gather*}
The root systems of the series $\mathbf{B}_n$ and $\mathbf{D}_n$
consist of the roots
\begin{gather*}
\pm e_i\pm e_j,\quad e_i,\quad i\neq j,\qquad\mbox{for}\quad\mathbf{B}_n,\\
\pm e_i\pm e_j ,\quad i\neq j,\qquad\mbox{for}\quad\mathbf{D}_n,
\end{gather*}
where $\{e_i\}^n_{i=1}$ stands for the standard orthonormal basis
in $\mathbb{C}^n$. The positive roots are as follows
\begin{gather*}
e_i\pm e_j,\quad i<j,\qquad e_i,\qquad\mbox{for}\quad\mathbf{B}_n,\\
e_i\pm e_j,\quad i<j,\qquad\mbox{for}\quad\mathbf{D}_n.
\end{gather*}
The set of all simple roots reads
\begin{gather*}
\alpha_k=e_k-e_{k+1},\qquad k=1,\ldots,n-1,\qquad \alpha_n=e_n \qquad\mbox{for}\quad\mathbf{B}_n,\\
\alpha_k=e_k-e_{k+1},\qquad\alpha_n=e_{n-1}+e_n,\qquad\mbox{for}\quad\mathbf{D}_n.
\end{gather*}

Let us illustrate these general results by a physical example
related to the ${\bf B}_2$ algebra. This algebra has two simple
roots $\alpha_1=e_1-e_2$, $\alpha _2=e_2 $, and two more positive
roots: $\alpha _1+\alpha _2=e_1 $ and $ \alpha _1+2\alpha _2=
e_1+e_2=\alpha _{\rm max} $. When they come as indices, e.g.\ in
$Q_\alpha(x,t)$ we will replace them by sequences of two positive
integers: $\alpha \to kn $ if $\alpha =k\alpha _1 + n\alpha _2 $;
if $\alpha =-(k\alpha _1 + n\alpha _2) $ we will use
$\overline{kn} $. The reduction which extracts the real forms of
${\bf B}_2 $ is
\begin{gather*}
K_1U^{\dag}(x,t,\lambda^*)K^{-1}_1 =U(x,t,\lambda ),
\end{gather*}
where $K_1 $ is an element of the Cartan subgroup:
$K_1=\mbox{diag} \, (s_1, s_2, 1, s_2, s_1) $ with $s_k=\pm 1 $,
$k=1,2 $.  This means that $J_i=J_i^* $, $i=1,2 $ and
$Q_\alpha(x,t)$ must satisfy:
\begin{gather*}
Q_{\overline{10}}(x,t)=-s_1s_2Q^{\ast}_{10}(x,t),
\qquad Q_{\overline{11}}(x,t)=-s_1Q^{\ast}_{11}(x,t),\\
Q_{\overline{12}}(x,t)=-s_1s_2Q^{\ast}_{12}(x,t), \qquad
Q_{\overline{01}}(x,t)=-s_2Q^{\ast}_{01}(x,t).
\end{gather*}
 and the
matrix $Q(x,t)$ is def\/ined by
\begin{gather*}
Q(x,t)  =  Q_{10}(x,t)E_{e_1-e_2}+Q_{12}(x,t)E_{e_1+e_2}+
Q_{11}(x,t)E_{e_1}+Q_{01}(x,t)E_{e_2}\\
\phantom{Q(x,t)  =}{}+ Q_{\overline{10}}(x,t)E_{-(e_1-e_2)}
+Q_{\overline{12}}(x,t)E_{-(e_1+e_2)}+Q_{\overline{11}}(x,t)E_{-e_1}
+Q_{\overline{01}}(x,t)E_{-e_2}.
\end{gather*}
As a result we get the following 4-wave system
\[i(J_1-J_2)Q_{10,t}(x,t)-i(I_1-I_2)Q_{10,x}(x,t)-ks_2Q_{11}(x,t)Q^{\ast}_{01}(x,t)=0,\]
\[iJ_1Q_{11,t}(x,t)-iI_1Q_{11,x}(x,t)-k(Q_{10}Q_{01}+s_2Q_{12}Q^{\ast}_{01})(x,t)=0,\]
\[i(J_1+J_2)Q_{12,t}(x,t)-i(I_1+I_2)Q_{12,x}(x,t)-kQ_{11}(x,t)Q_{01}(x,t)=0,\]
\[iJ_2Q_{01,t}(x,t)-iI_2Q_{01,x}(x,t)-ks_1(Q^{\ast}_{11}Q_{12}+s_2Q^{\ast}_{10}Q_{11})(x,t)=0,\]
where $k=J_1I_2-J_2I_1$ and its Hamiltonian $H(t)=H_0(t)+H_{\rm
int}(t) $ is
\begin{gather*}
H_0(t)={i\over 2}\int_{-\infty }^{\infty }dx \, \big[(I_1-I_2)
(Q_{10}(x,t)Q_{10,x}^*(x,t)-Q_{10,x}(x,t)Q_{10}^*(x,t))\\
 \phantom{H_0(t)=}{}+I_2(Q_{01}(x,t)Q_{01,x}^*(x,t)-Q_{01,x}(x,t)Q_{01}^*(x,t))
+I_1(Q_{11}(x,t)Q_{11,x}^*(x,t)\nonumber \\
\phantom{H_0(t)=}{}-Q_{11,x}(x,t)Q_{11}^*(x,t))+(I_1+I_2)
(Q_{12}(x,t)Q_{12,x}^*(x,t)
- Q_{12,x}(x,t)Q_{12}^*(x,t)) \big],\nonumber\\
H_{{\rm int}}(t)= 2k s_1\int_{-\infty }^{\infty }dx\, \left[s_2
(Q_{12}(x,t)Q_{11}^*(x,t)Q_{01}^*(x,t)+Q_{12}^*(x,t)Q_{11}(x,t)Q_{01}(x,t))
\right.\nonumber\\
\left. \phantom{H_{{\rm
int}}(t)=}{}+(Q_{11}(x,t)Q_{01}^*(x,t)Q_{10}^*(x,t)+Q_{11}^*(x,t)Q_{01}(x,t)Q_{10}(x,t))\right],
\nonumber
\end{gather*}
and the symplectic form:
\begin{gather*}
\Omega ^{(0)}  =  i \int_{-\infty }^{\infty } dx \, \big[
(J_1-J_2)\delta Q_{10}(x,t)\wedge\delta Q_{10}^*(x,t) + J_2\delta
Q_{01}(x,t)
\wedge \delta Q_{01}^*(x,t)\nonumber\\
\phantom{\Omega ^{(0)}  =}{} +J_1\delta Q_{11}(x,t)\wedge \delta
Q_{11}^*(x,t)+ (J_1+J_2)\delta Q_{12}(x,t)
\wedge \delta Q_{12}^*(x,t)\big].
\end{gather*}

The invariance condition (\ref{invariance}) leads to the following
form of the dressing factor
\begin{gather}\label{factor1}
g(x,t,\lambda)=\openone+\frac{A(x,t)}{\lambda-\lambda^{+}}
+\frac{K_1SA^{\ast}(x,t)(K_1S)^{-1}}{\lambda-(\lambda^{+})^{\ast}},
\end{gather}
i.e.\ comparing with (\ref{g}) we see that
\[
B(x,t)=K_1SA^{\ast}(x,t)(K_1S)^{-1},\qquad
\lambda^{-}=(\lambda^{+})^{\ast}.
\]
By taking the limit $\lambda\to\infty$ in equation (\ref{factor
PDE}) and taking into account the explicit formula (\ref{factor1})
one can derive the following relation between the bare solution
$Q_0(x,t)$ and the dressed one $Q(x,t)$
\begin{gather*}
[J,Q(x,t)]=[J,Q_{0}(x,t)+A(x,t)+K_1SA^{\ast}(x,t)SK_1].
\end{gather*}
Thus the one soliton solution of the $N$-wave equation is
determined by  one matrix-valued function $A(x,t)$. We will obtain
$A(x,t)$ in two steps by deriving certain algebraic and
dif\/ferential relations. First of all recall that the dressing
factor $g(x,t,\lambda)$ must belong to the orthogonal group
$\mathrm{SO}(n,\mathbb{C})$, hence
\[
g^{-1}(x,t,\lambda)=S^{-1}g^T(x,t,\lambda)S.
\]
Besides, the equality $gg^{-1}=1$ must hold identically with
respect to $\lambda$, therefore $A(x,t)$ satisf\/ies the algebraic
restrictions
\begin{gather}\label{algrestr}
A(x,t)SA^T(x,t)=0,\qquad A(x,t)S\omega^T(x,t)+\omega(x,t)
SA^T(x,t)=0,
\end{gather}
where
\[
\omega(x,t)=\openone+\frac{K_1SA^{\ast}(x,t)SK_1}{2i\nu},\qquad
\lambda^{+}=\mu+i\nu.
\]
From the f\/irst equality in (\ref{algrestr}) it follows that the matrix $A(x,t)$ is a
degenerate one and it can be decomposed
\[
A(x,t)=X(x,t)F^T(x,t),
\]
where $X(x,t)$ and $F(x,t)$ are $n\times k$ ($1\leq k<n$) matrices
of maximal rank $k$. The equalities~(\ref{algrestr}) can be
rewritten in terms of $X(x,t)$ and $F(x,t)$ as follows
\[
F^T(x,t)SF(x,t)=0,\qquad X(x,t)F^T(x,t)S\omega^T(x,t)+\omega(x,t)
SFX^T(x,t)=0
\]
or introducing a $k\times k$ skew symmetric matrix $\alpha(x,t)$
the latter restriction reads
\begin{gather}\label{algsys}
\left(\openone+\frac{K_1SX^{\ast}(x,t)F^{\dag}(x,t)SK_1}{2i\nu}\right)SF(x,t)
=X(x,t)\alpha(x,t).
\end{gather}

Another type of restrictions concerning the matrix-valued
functions $F(x,t)$ and $\alpha(x,t)$ comes form the
$\lambda$-independence of the potential $[J,Q(x,t)]$. If we
express the potential in the equation~(\ref{factor PDE}) we get
\begin{gather}\label{independent}
[J,Q(x,t)]=-i\partial_x g
g^{-1}(x,t,\lambda)+g[J,Q_0(x,t)]g^{-1}(x,t,\lambda)
+\lambda\left(J-gJg^{-1}(x,t,\lambda)\right).
\end{gather}
Annihilation of residues in (\ref{independent}) leads to the
following linear dif\/ferential equations
\[
i\partial_xF^T(x,t)-F^T(x,t)([J,Q_0(x,t)]-\lambda^{+}J)=0
\]
and
\[
i\partial_x\alpha(x,t)+F^T(x,t)JSF(x,t)=0.
\]
After integration we obtain that
\[
F(x,t)=S\chi^{+}_0(x,t,\lambda^{+})SF_0,
\]
where the constant matrix $F_0$ obeys the following equality
\[
F^T_0SF_0=0
\]
and the matrix $\alpha(x,t)$ reads
\[
\alpha(x,t)=F^T_0(\chi^{+}_0(x,t,\lambda^{+}))^{-1}\partial_{\lambda}{\chi}^{+}_0(x,t,
\lambda^{+})SF_0+\alpha_0.
\] In the soliton case the fundamental
solution is just a plane wave. Therefore functions $F(x,t)$ and
$\alpha(x,t)$ get the form
\begin{gather*}
F(x,t)=e^{i\lambda^{+}(Jx+It)}F_0,\qquad\alpha(x,t)= i
F^T_0(Jx+It)SF_0 + \alpha_0.
\end{gather*}
It remains to f\/ind the factor $X(x,t)$. For this purpose we
solve the linear equation (\ref{algsys}) and f\/ind that
\[X(x,t)=2\nu(iK_1F^{\ast}-2\nu SF(F^{\dag}K_1F)^{-1}\alpha^{\ast})\big(F^TK_1F^{\ast}-4\nu^2\alpha (F^{\dag}K_1F)^{-1}\alpha^{\ast}\big)^{-1}.\]
In particular, in the simplest case when
$\mbox{rank}X(x,t)=\mbox{rank}F(x,t)=1$ we have $\alpha(x,t)\equiv
0$ and therefore
\[
X(x,t)=\frac{2i\nu}{F^{\dag}(x,t)K_1F(x,t)}K_1F^{\ast}(x,t).
\]
The soliton solution obtains the form
\[
Q_{ij}(x,t)=\left\{\begin{array}{ll} \displaystyle
\frac{2i\nu}{F^{\dag}(x,t)K_1F(x,t)}\big((K_1F^{\ast}(x,t))_iF_j(x,t)
\vspace{1mm}\\
\qquad{}+(-1)^{i+j+1}F_{6-i}(x,t))(K_1F^{\ast}(x,t))_{6-j}\big)
 ,& i\neq j, \vspace{1mm}\\ 0, & i=j. \end{array}\right.
 \]
Turning back to the $\mathbf{B}_2$ case this result can be written
by components as follows
\begin{gather*} Q_{10}(x,t) = \frac{2i\nu}{F^{\dag}K_1F}\left(s_1F^{\ast}_{0,1}F_{0,2}e^{i\left(\lambda^{+}z_2
-{\lambda^{+}}^{\ast}z_1\right)}
+s_2F^{\ast}_{0,4}F_{0,5}e^{i\left({\lambda^{+}}^{\ast}z_2-\lambda^{+}z_1\right)}\right),\\
 Q_{11}(x,t) = \frac{2i\nu}{F^{\dag}K_1F}\left(s_1F^{\ast}_{0,1}F_{0,3}e^{-i{\lambda^{+}}^{\ast}z_1}
-F^{\ast}_{0,3}F_{0,5}e^{-i\lambda^{+} z_1}\right), \\
 Q_{12}(x,t) =  \frac{2i\nu}{F^{\dag}K_1F}\left(s_1F^{\ast}_{0,1}F_{0,4}e^{-i\left({\lambda^{+}}^{\ast}z_1
+\lambda^{+}z_2\right)}
+s_2F^{\ast}_{0,2}F_{0,5}e^{-i\left(\lambda^{+}z_1+{\lambda^{+}}^{\ast}z_2\right)}\right),\\
Q_{01}(x,t)=\frac{2i\nu}{F^{\dag}K_1F}\left(s_2F^{\ast}_{0,2}F_{0,3}e^{-i{\lambda^{+}}^{\ast}z_2}
+F^{\ast}_{0,3}F_{0,4}e^{-i\lambda^{+}z_2}\right),\\
F^{\dag}K_1F=s_1|F_{0,1}|^2e^{-2\nu z_1}+s_2|F_{0,2}|^2e^{-2\nu
z_2}+|F_{0,3}|^3
+s_2|F_{0,4}|^2e^{2\nu z_2}+s_1|F_{0,5}|^2e^{2\nu z_1},\\
 z_{\sigma}=J_{\sigma}x+I_{\sigma}t,\qquad \sigma=1,2.
 \end{gather*}

A natural choice for the matrix which determines the action of the
reduction group $G_R$ is $K_1=\openone$. Let us also require the
symmetry conditions
\[F_{0,1}=F_{0,5}^{\ast},\qquad F_{0,2}=F_{0,4}^{\ast},\qquad F_{0,3}=F_{0,3}^{\ast}.\]
As a result the one-soliton solution simplif\/ies signif\/icantly
\begin{gather*}
Q_{10}(x,t)=\frac{i\nu}{\Delta_1}\sinh 2\theta_0\cosh(\nu(z_1+z_2))e^{-i\mu(z_1-z_2+\phi_1-\phi_2)},\\
Q_{11}(x,t)=-\frac{2\sqrt{2}i\nu}{\Delta_1}\sinh\theta_0\sinh(\nu z_1)e^{-i(\mu z_1+\phi_1)},\\
Q_{12}(x,t)=\frac{i\nu}{\Delta_1}\sinh(2\theta_0)\cosh(\nu(z_1-z_2))e^{-i\mu(z_1+z_2+\phi_1+\phi_2)},\\
Q_{01}(x,t)=\frac{2\sqrt{2}i\nu}{\Delta_1}\cosh\theta_0\cosh(\nu
z_2)e^{-i\mu(z_2+\phi_2)},
\end{gather*}
where we have used the representation
\begin{gather*}
F_{0,1}=\frac{F_{0,3}}{\sqrt{2}}\sinh\theta_0 e^{i\phi_1},\qquad
F_{0,2}=\frac{F_{0,3}}{\sqrt{2}}\cosh\theta_0 e^{i\phi_2},\\
\Delta_1(x,t)=2\left(\sinh^2\theta_0\sinh^2(\nu
z_1)+\cosh^2\theta_0\cosh^2(\nu z_2)\right),
\end{gather*}
where $\theta_{0}$ is an arbitrary real constant.

If we apply this dressing procedure to the one-soliton solution we
obtain a two-soliton solution. Iterating this process we can
generate multisoliton solutions, i.e.
\[0\stackrel{g_{\rm 1s}}{\longrightarrow} Q_{\rm 1s}(x,t)\stackrel{g_{\rm 2s}}{
\longrightarrow}Q_{\rm 2s}(x,t)\longrightarrow\cdots
\stackrel{g_{m \rm s}}{\longrightarrow}Q_{m\rm s}(x,t).\]

The particular case $s_1=s_2=1 $ leads to the compact real form
$\mathfrak{so}(5,0)\simeq\mathfrak{so}(5,\bbbr) $ of the ${\bf
B}_2 $-algebra. The choice $s_1=-s_2=-1 $ leads to the noncompact
real form $\mathfrak{so}(2,3) $ and $s_1=s_2=-1 $ gives another
noncompact one -- the Lorentz algebra $\mathfrak{so}(1,4)$. If
$s_{1}=s_{2}=1$ and we identify
\begin{gather*} 
 Q_{01}(x,t)=-\frac{i}{\kappa}Q_{\rm pol}(x,t), \qquad Q_{10}(x,t)=- \frac{i}{\kappa}
E_{\rm s}(x,t),\nonumber\\
Q_{11}(x,t)=-\frac{i}{\kappa} E_{\rm p}(x,t), \qquad
Q_{12}(x,t)=-\frac{i}{\kappa} E_{\rm a}(x,t), \nonumber
\end{gather*}
then we obtain the system similar to that studied in
\cite{AckMil,gk96} which describes Stokes-anti-Stokes wave
generation. Here $Q_{\rm pol}(x,t) $ is the normalized ef\/fective
polarization of the medium and $E_{\rm p}(x,t) $, $E_{\rm s}(x,t)
$ and $E_{\rm a}(x,t) $ are the normalized pump, Stokes and
anti-Stokes wave amplitudes respectively. In this case the matrix
$Q(x,t)$ is def\/ined by
\begin{gather*}
Q(x,t) =-\frac{i}{\kappa}\left(\begin{array}{ccccc}
0 & E_{\rm s}  & E_{\rm p} & E_{\rm a} & 0 \\
E_{\rm s}^{*} & 0 & Q_{\rm pol} & 0 & E_{\rm a} \\
E_{\rm p}^{*} & Q_{\rm pol}^{*} & 0 & Q_{\rm pol} &
-E_{\rm p}\\
E_{\rm a}^{*} & 0 & Q_{\rm pol}^{*} & 0 & E_{\rm s} \\
0 & E_{\rm a}^{*} &  -E_{\rm p}^{*} & E_{\rm s}^{*} & 0
\end{array} \right), \nonumber
\end{gather*}
and for 4-wave system we have
\begin{gather*}\label{eq:b2.4b}
(J_1-J_2) E_{{\rm s},t}(x,t)-(I_1-I_2) E_{{\rm s},x}(x,t)- E_{{\rm
p}}(x,t) Q_{{\rm pol}}^{*}(x,t)=0,
\nonumber\\
J_2 Q_{{\rm pol},t}(x,t)-I_2 Q_{{\rm pol},x}(x,t)-E_{{\rm
p}}^{*}(x,t) E_{{\rm a}}(x,t) - E_{{\rm p}}(x,t)E_{{\rm
s}}^{*}(x,t)=0,
\nonumber\\
J_1 E_{{\rm p},t}(x,t)-I_1 E_{{\rm p},x}(x,t)-  E_{{\rm a}}(x,t)
Q_{{\rm pol}}^{*}(x,t)+E_{{\rm s}}(x,t)Q_{{\rm pol}}(x,t)=0,
\nonumber\\
(J_1+J_2)E_{{\rm a},t}(x,t)-(I_1+I_2)E_{{\rm a},x}(x,t)+ E_{\rm
p}(x,t) Q_{\rm pol}(x,t)=0. \nonumber
\end{gather*}
Finally we can rewrite it in the form:
\begin{gather}
\frac{1}{v_{-1}} \frac{\partial E_{\rm s}(x,t)}{\partial t}+
\frac{\partial E_{\rm s}(x,t)}{\partial x}- \kappa_{-1}E_{{\rm
p}}(x,t)
Q_{{\rm pol}}^{*}(x,t)=0, \nonumber\\
\frac{\partial Q_{\rm pol}(x,t)}{\partial t}-\kappa_{pol}(E_{{\rm
p}}^{*}(x,t) E_{{\rm a}}(x,t) + E_{{\rm p}}(x,t)E_{{\rm
s}}^{*}(x,t))=0,
\nonumber\\
\frac{1}{v_{0}}  \frac{\partial E_{\rm p}(x,t)}{\partial t}+
\frac{\partial E_{\rm p}(x,t)}{\partial x}-\kappa_{0}( E_{{\rm
a}}(x,t)
Q_{{\rm pol}}^{*}(x,t)-E_{{\rm s}}(x,t) Q_{{\rm pol}}(x,t))=0,\label{eq:b2.4s} \\
\frac{1}{v_{1}} \frac{\partial E_{\rm a}(x,t)}{\partial t}+
\frac{\partial E_{\rm a}(x,t)}{\partial x}+\kappa_{1}E_{\rm
p}(x,t) Q_{\rm pol}(x,t)=0, \nonumber
\end{gather}
where
\begin{gather*}
v_{-1}=\frac{d\omega_{-1}}{d
k_{-1}}=-\frac{I_{1}-I_{2}}{J_{1}-J_{2}}, \qquad
\kappa_{-1}=-\frac{1}{I_{1}-I_{2}}, \\
v_{0}=\frac{d\omega_{0}}{d k_{0}}=-\frac{I_{1}}{J_{1}},
\qquad \kappa_{0}=-\frac{1}{I_{1}}, \qquad I_2=0, \\
v_{1}=\frac{d\omega_{1}}{d
k_{1}}=-\frac{I_{1}+I_{2}}{J_{1}+J_{2}}, \qquad
\kappa_{1}=-\frac{1}{I_{1}+I_{2}},\qquad\kappa_{\rm
pol}=\frac{1}{J_{2}}.
\end{gather*}
The particular case $s_1=s_2=1$, $I_2=0 $, the one-soliton
solution obeys (\ref{eq:b2.4s}). Finally for the simplest one
soliton solution of (\ref{eq:b2.4s})  the  modulus squared  of one
soliton amplitudes are given~by
\begin{gather*}
|E_{\rm{s}}(x,t)|^2=\frac{\kappa^2\nu^2}{\Delta_1^2}\sinh^2 2\theta_0\cosh^2 \nu[(J_1+J_2)x+I_1t],\\
|E_{\rm{p}}(x,t)|^2=\frac{8\kappa^2\nu^2}{\Delta_1^2}\sinh^2\theta_0\sinh^2 \nu(J_1x+I_1t),\\
|E_{\rm{a}}(x,t)|^2=\frac{\kappa^2\nu^2}{\Delta_1^2}\sinh^2 2\theta_0\cosh^2 \nu[(J_1-J_2)x+I_1t],\\
|Q_{\rm{pol}}(x,t)|^2=\frac{8\kappa^2\nu^2}{\Delta_1^2}\cosh^2\theta_0\cosh^2
\nu J_2x,
\end{gather*}
where
\[
\kappa=-J_2I_1,\qquad\Delta_1(x,t)=2\left(\sinh^2\theta_0\sinh^2(\nu
(J_1x+I_1t)) +\cosh^2\theta_0\cosh^2(\nu J_2x)\right).
\]
 Note that the
canonical reduction ensures that $\Delta_1(x,t)$ is positive for
all $x$ and $t$, so the solitons of this model can have no
singularities.

\subsection[Another type of $\mathbb{Z}_2$ reduction]{Another type of $\boldsymbol{\mathbb{Z}_2}$ reduction}

Let us consider the $\mathbb{Z}_2$ reduction
\[\chi^{-}(x,t,\lambda)=K_2\left\{[\chi^{+}(x,t,-\lambda)]^T\right\}^{-1}K_2^{-1}.\]
where $K_2=\diag(s_1,s_2,1,s_4,s_5)$ and $s_1=s_5=\pm 1$,
$s_2=s_4=\pm 1$.

Therefore we have the symmetry conditions
\[K_2U(x,t,-\lambda)^TK_2^{-1}=-U(x,t,\lambda)
\quad\Rightarrow \quad Q(x,t)=K_2Q^T(x,t)K_2^{-1}.\] In particular
if we choose $K_2=\openone$ then $Q(x,t)$ is a symmetric matrix.

The invariance condition implies that the dressing matrix gets the
form
\[g(x,t,\lambda)=\openone+\frac{A(x,t)}{\lambda-\lambda^{+}}-
\frac{K_2SA(x,t)(K_2S)^{-1}}{\lambda+\lambda^{+}}.\] Hence here
the poles of the dressing factor form a doublet $\{\lambda^+,
-\lambda^+\}$ whose residues are related~by:
\[B(x,t)=-K_2SA(x,t)(K_2S)^{-1},\qquad \lambda^{-}=-\lambda^{+}.\]
The soliton solution is expressed by the residues of the dressing
matrix as follows
\[[J,Q(x,t)]=[J,A(x,t)-K_2SA(x,t)SK_2].\]
Like before one can present the matrix $A(x,t)$ in the following
way
\[A(x,t)=X(x,t)F^T(x,t),\]
where $F(x,t)=S\chi^{+}_0(x,t,\lambda^{+})SF_0$. In the soliton
case it is just the well known plane wave $F(x,t)=e^{i\lambda^{+}
(Jx+It)}F_0$. The other factor $X(x,t)$ is given by
\[
X(x,t)=2\lambda^{+}\left(K_2F-2\lambda^{+}SF(F^TK_2F)^{-1}\alpha\right)\left(F^TK_2F
-4(\lambda^{+})^2\alpha(F^TK_2F)^{-1}\alpha\right)^{-1}.
\] As a
particular case when $\mbox{rank}X(x,t)=\mbox{rank}F(x,t)=1$ and
$\alpha(x,t)\equiv 0$ it is simply
\[
X(x,t)=\frac{2\lambda^{+}K_2F(x,t)}{F^T(x,t)K_2F(x,t)}, \qquad
F^T(x,t)K_2F(x,t)=\sum^5_{k=1}s_{k}
e^{2i\lambda^{+}(J_kx+I_kt)}F^2_{0,k}.
\] Consequently the
one-soliton solution reads
\begin{gather}\label{eq:no}
Q_{ij}(x,t)=\left\{\begin{array}{ll} \displaystyle
\frac{2\lambda^{+}}{F^T(x,t)K_2F(x,t)}\left((K_2F)_iF_j+
(-1)^{i+j+1}F_{6-i}(K_2F)_{6-j}\right),& i\neq j,\vspace{1mm}\\
0 ,&i=j.\end{array}\right.
\end{gather}

In the case of $\mathbf{B}_2$ algebra there are only 4 independent
f\/ields as shown below
\[
Q(x,t)=\left(\begin{array}{ccccc}
0                 & Q_{10}(x,t)       & Q_{11}(x,t)     & Q_{12}(x,t)       & 0          \\
s_1s_2Q_{10}(x,t) & 0                 & Q_{01}(x,t)     & 0                 & Q_{12}(x,t) \\
s_1Q_{11}(x,t)    & s_2Q_{01}(x,t)    & 0               & Q_{01}(x,t)       & -Q_{11}(x,t)\\
s_1s_2Q_{12}(x,t) & 0                 & s_2Q_{01}(x,t)  & 0                 & Q_{10}(x,t) \\
0                 & s_1s_2Q_{12}(x,t) & -s_1Q_{11}(x,t) &
s_1s_2Q_{10}(x,t) & 0
\end{array}\right).
\]

The corresponding 4-wave system reads
\begin{gather*}
 i(J_1-J_2)Q_{10,t}(x,t)-i(I_1-I_2)Q_{10,x}(x,t)+ks_2Q_{11}(x,t)Q_{01}(x,t)=0,\\
i J_1Q_{11,t}(x,t)-i I_1Q_{11,x}(x,t)+kQ_{01}(x,t)(s_2Q_{12}(x,t)-Q_{10}(x,t))=0,\\
i(J_1+J_2)Q_{12,t}(x,t)-i(I_1+I_2)Q_{12,x}(x,t)-kQ_{11}(x,t)Q_{01}(x,t)=0,\\
i J_2Q_{01,t}(x,t)-i
I_2Q_{01,x}(x,t)+ks_1Q_{11}(x,t)(Q_{12}(x,t)+s_2Q_{10}(x,t))=0.
\end{gather*}

Its one-soliton solution when $K_2=\openone$ is presented by the
expressions
\begin{gather*}
Q_{10}(x,t)=\frac{2\lambda^{+}}{F^TF}\left(e^{i\lambda^{+}[(J_1+J_2)x+(I_1+I_2)t]}F_{0,1}F_{0,2}
+e^{-i\lambda^{+}[(J_1+J_2)x+(I_1+I_2)t]}F_{0,4}F_{0,5}\right),\\
Q_{11}(x,t)=\frac{2\lambda^{+}F_{0,3}}{F^TF}\left(e^{i\lambda^{+}(J_1x+I_1t)}F_{0,1}
-e^{-i\lambda^{+}(J_1x+I_1t)}F_{0,5}\right),\\
Q_{12}(x,t)=\frac{2\lambda^{+}}{F^TF}\left(e^{i\lambda^{+}[(J_1-J_2)x+(I_1-I_2)t]}F_{0,1}F_{0,4}
+e^{-i\lambda^{+}[(J_1-J_2)x+(I_1-I_2)t]}F_{0,2}F_{0,5}\right),\\
Q_{01}(x,t)=\frac{2\lambda^{+}F_{0,3}}{F^TF}\left(e^{i\lambda^{+}(J_2x+I_2t)}F_{0,2}
+e^{-i\lambda^{+}(J_2x+I_2t)}F_{0,4}\right),
\end{gather*}
 where
$F^T(x,t)F(x,t)$ is obtained from the expression in (\ref{eq:no})
putting $s_k=1$. After imposing the additional restriction
$F_{0,1}=F_{0,5}$ and $F_{0,2}=F_{0,4}$ we get
\begin{gather*}
Q_{10}(x,t)= \frac{\lambda^{+}}{\Delta_2(x,t)}\sinh
2\theta_0\cos\lambda^{+}[(J_1+J_2)x+(I_1+I_2)t],\nonumber \\
Q_{11}(x,t)=\frac{2\sqrt{2}\,i\lambda^{+}}{\Delta_2(x,t)}\sinh
2\theta_0\sin\lambda^{+}(J_1x+I_1t),\\ 
Q_{12}(x,t)=\frac{\lambda^{+}}{\Delta_2(x,t)}\sinh
2\theta_0\cos\lambda^{+}[(J_1-J_2)x+(I_1-I_2)t],\nonumber \\
Q_{01}(x,t)=\frac{2\sqrt{2}\lambda^{+}}{\Delta_2(x,t)}\cosh
2\theta_0\cos\lambda^{+}(J_2x+I_2t),\nonumber
\end{gather*}
where
\begin{gather}\label{Repr_F12}
F_{0,1}=\frac{F_{0,3}}{\sqrt{2}}\sinh\theta_0 ,\qquad
F_{0,2}=\frac{F_{0,3}}{\sqrt{2}}\cosh\theta_0,
\end{gather}
$\theta_0$ is a complex parameter and
\begin{gather*}
\Delta_2(x,t)=2(\cosh^2\theta_0\cos^2\lambda^{+}(J_2x+I_2t)-
\sinh^2\theta_0\sin^2\lambda^{+}(J_1x+I_1t)).
\end{gather*}

In its turn the existence of such reduction leads to the existence
of a special class of solitons, the so-called breathers, in the
case when there are two $\mathbb{Z}_2$ reductions (canonical one
and another one of the type mentioned above) applied to the
$N$-wave systems.

\section[$N$-wave equations with a $\mathbb{Z}_2\times\mathbb{Z}_2$-reduction
and their soliton solutions]{$\boldsymbol{N}$-wave equations with
a $\boldsymbol{\mathbb{Z}_2\times\mathbb{Z}_2}$-reduction\\ and
their soliton solutions}\label{sec:3}

In this section we consider a
$\mathbb{Z}_2\times\mathbb{Z}_2$-reduced $N$-wave system
associated to the orthogonal algebra. Such systems admit real
valued solutions. The soliton solutions can be classif\/ied into
two major types: ``doublet'' solitons which are related to two
imaginary discrete eigenvalues (a~``doublet'') of $L$ and solitons
which correspond to four eigenvalues (a ``quadruplet'').

\subsection{Doublet solitons}

Let the action of $\mathbb{Z}_2\times\mathbb{Z}_2$ in the space of
fundamental solutions of the linear problem is given by
\begin{gather*}
\chi^{-}(x,t,\lambda)=K_1\left((\chi^{+})^{\dag}(x,t,\lambda^{\ast})\right)^{-1}K^{-1}_1,\\
\chi^{-}(x,t,\lambda)=K_2\left((\chi^{+})^T(x,t,-\lambda)\right)^{-1}K^{-1}_2,
\end{gather*}
where $K_{1,2}\in SO(n)$ and $[K_1,K_2]=0$. Consequently the
potential $U(x,t,\lambda)$ satisf\/ies the following symmetry
conditions
\begin{gather*}
K_1U^{\dag}(x,t,\lambda^{\ast})K^{-1}_1=U(x,t,\lambda),\qquad K_1J^{\ast}K^{-1}_1=J,\\
K_2U^T(x,t,-\lambda)K^{-1}_2=-U(x,t,\lambda),\qquad
K_2JK^{-1}_2=J.
\end{gather*}
In accordance with what we said in previous chapter the dressing
factor $g(x,t,\lambda)$ must be invariant under the action of
$\mathbb{Z}_2\times\mathbb{Z}_2$, i.e.
\begin{gather}\label{ginv_1}
K_1\left(g^{\dag}(x,t,\lambda^{\ast})\right)^{-1}K^{-1}_1=g(x,t,\lambda),
\\
\label{ginv_2}
K_2\left(g^T(x,t,-\lambda)\right)^{-1}K^{-1}_2=g(x,t,\lambda).
\end{gather}
Let for the sake of simplicity require that $K_1=K_2=\openone$. As
a result we f\/ind that the poles of the dressing matrix can be
purely imaginary, i.e.
\[\lambda^{\pm}=\pm i\nu,\qquad \nu>0 .\]
Thus the invariance condition implies that the dressing matrix
gets the form
\[g(x,t,\lambda)=\openone+\frac{A(x,t)}{\lambda-i\nu}+\frac{SA^{\ast}(x,t)S}{\lambda+i\nu}, \qquad A^{\ast}(x,t)=-A(x,t).\]
Following already discussed procedures we derive in the simplest
case that the explicit form of $A(x,t)$ is
\[A(x,t)=\frac{2i\nu}{F^T(x,t)F(x,t)}F(x,t)F^T(x,t),\]
where the vector $F(x,t)=e^{-\nu(Jx+It)}F_0$ is real. Consequently
the soliton solution written in a~standard matrix notation is
\begin{gather}\label{doublon}
Q_{kl}(x,t)=\left\{\begin{array}{ll}
0 ,& \mbox{if}\ k=l, \vspace{1mm}\\
\displaystyle \frac{2i\nu}{F^T(x,t)F(x,t)}\big(F_k(x,t)F_l(x,t)\vspace{1mm}\\
\qquad{}+(-1)^{k+l+1}F_{6-k}(x,t)F_{6-l}(x,t)\big) ,& \mbox{if}\
k\neq l.
\end{array}\right.
\end{gather}

In the case of $\mathbf{B}_2$ algebra we obtain the following
$\mathbb{Z}_2\times\mathbb{Z}_2$-reduced 4-wave system
\begin{gather*}
(J_1-J_2)q_{10,t}(x,t)-(I_1-I_2)q_{10,x}(x,t)+kq_{11}(x,t)q_{01}(x,t)=0,\\
J_1q_{11,t}(x,t)-I_1q_{11,x}(x,t)+k(q_{12}(x,t)-q_{10}(x,t))q_{01}(x,t)=0,\\
(J_1+J_2)q_{12,t}(x,t)-(I_1+I_2)q_{12,x}(x,t)-kq_{11}(x,t)q_{01}(x,t)=0,\\
J_2q_{01,t}(x,t)-
I_2q_{01,x}(x,t)+k(q_{10}(x,t)+q_{12}(x,t))q_{11}(x,t)=0,
\end{gather*}
where $q_{10}(x,t)$, $q_{11}(x,t)$, $q_{12}(x,t)$ and
$q_{01}(x,t)$ are real valued f\/ields and their indices are
associated with the basis of simple roots of $\mathbf{B}_2$
introduced in the previous section, i.e.
\begin{gather*}
Q_{10}(x,t)=i q_{10}(x,t),\qquad Q_{11}(x,t)=i q_{11}(x,t),\\
 Q_{12}(x,t)=i q_{12}(x,t),\qquad Q_{01}(x,t)=i
q_{01}(x,t).
\end{gather*}
The ``bonding'' constant $k$ coincides with the one in the
previous examples
\[
k=J_1I_2-J_2I_1.
\]
Therefore the solution (\ref{doublon}) turns into
\begin{gather*}
q_{10}(x,t)=\frac{2\nu}{F^TF}\left(e^{-\nu[(J_1+J_2)x+(I_1+I_2)t]}F_{0,1}F_{0,2}
+e^{\nu[(J_1+J_2)x+(I_1+I_2)t]}F_{0,5}F_{0,4}\right),\\
q_{11}(x,t)=\frac{2\nu}{F^TF}\left(e^{-\nu(J_1x+I_1t)}F_{0,1}F_{0,3}
-e^{\nu(J_1x+I_1t)}F_{0,5}F_{0,3}\right),\\
q_{12}(x,t)=\frac{2\nu}{F^TF}\left(e^{-\nu[(J_1-J_2)x+(I_1-I_2)t]}F_{0,1}F_{0,4}
+e^{\nu[(J_1-J_2)x+(I_1-I_2)x]}F_{0,5}F_{0,2}\right),\\
q_{01}(x,t)=\frac{2\nu}{F^TF}\left(e^{-\nu(J_2x+I_2t)}F_{0,2}F_{0,3}
+e^{\nu(J_2x+I_2t)}F_{0,4}F_{0,3}\right).
\end{gather*}
These solutions can be rewritten in terms of hyperbolic functions
as follows
\begin{gather*}
q_{10}(x,t)=\frac{4\nu}{F^TF}N_1N_2\cosh \{\nu[(J_1+J_2)x+(I_1+I_2)t]+\delta_1+\delta_2\},\\
q_{11}(x,t)=-\frac{4\nu}{F^TF}N_1F_{0,3}\sinh[\nu(J_1x+I_1t)+\delta_1],\\
q_{12}(x,t)=\frac{4\nu}{F^TF}N_1N_2\cosh\{\nu[(J_1-J_2)x+(I_1-I_2)t]
+\delta_1-\delta_2\},\\
q_{01}(x,t)=\frac{4\nu}{F^TF}N_2F_{0,3}\cosh[\nu(J_2x+I_2t)+\delta_2],\\
F^T(x,t)F(x,t)=2N^2_1\cosh 2(\nu(J_1x+I_1t)+\delta_1)+2N^2_2\cosh
2(\nu(J_2x+I_2t)+\delta_2)+F^2_{0,3},
\end{gather*}
where we have implied that $F_{0,k}>0$ for $k=1,2,4,5$ and
therefore the following expressions
\[\delta_1=\frac{1}{2}\ln\frac{F_{0,5}}{F_{0,1}},\qquad\delta_2=\frac{1}{2}\ln\frac{F_{0,4}}{F_{0,2}},
\qquad N_1=\sqrt{F_{0,1}F_{0,5}},\qquad
N_2=\sqrt{F_{0,2}F_{0,4}}\] make sense.

In particular when $F_{0,1}=F_{0,5}$ and $F_{0,2}=F_{0,4}$ or in
other words $\delta_1=\delta_2=0$ we can apply the same
representation as in (\ref{Repr_F12}) and as a result we obtain
\begin{gather*}
q_{10}(x,t)=\frac{\nu}{\Delta_{D}}\sinh(2\theta_0)\cosh\nu[(J_1+J_2)x+(I_1+I_2)t],\\
q_{11}(x,t)=-\frac{2\sqrt{2}\nu}{\Delta_D}\sinh\theta_0\sinh\nu(J_1x+I_1t),\\
q_{12}(x,t)=\frac{\nu}{\Delta_D}\sinh(2\theta_0)\cosh\nu[(J_1-J_2)x+(I_1-I_2)t],\\
q_{01}(x,t)=\frac{2\sqrt{2}\nu}{\Delta_D}\cosh\theta_0\cosh\nu(J_2x+I_2t),
\end{gather*}
where $\theta_0\in\mathbb{R}$ and
\[
\Delta_{D}(x,t)=2\left(\sinh^2\theta_0\sinh^2\nu(J_1x+I_1t)+\cosh^2\theta_0\cosh^2\nu(J_2x+I_2t)\right).
\]

\subsection{Quadruplet solitons}\label{sec:3.1}

There is another way to ensure the
$\mathbb{Z}_2\times\mathbb{Z}_2$ invariance of the dressing
factor. This time we consider dressing factors  $g(x,t,\lambda)$
with two more poles. The requirements
(\ref{ginv_1})--(\ref{ginv_2}) lead to the following dressing
matrix
\begin{gather}
g(x,t,\lambda) = \openone+\frac{A(x,t)}{\lambda-\lambda^{+}}
+\frac{K_1SA^{\ast}(x,t)(K_1S)^{-1}}{\lambda-(\lambda^{+})^{\ast}}
 -\frac{K_2SA(x,t)(K_2S)^{-1}}{\lambda+\lambda^{+}}\nonumber\\
 \phantom{g(x,t,\lambda) =}{}-\frac{K_1K_2A^{\ast}(x,t)(K_1K_2)^{-1}}{\lambda+(\lambda^{+})^{\ast}}.\label{factor}
\end{gather}

By taking the limit $\lambda\to\infty$ in equation (\ref{factor
PDE}) and taking into account the explicit
formu\-la~(\ref{factor}) one can derive in the soliton case
$Q_0(x,t)\equiv 0$ the following relation
\begin{gather}\label{breather}
[J,Q](x,t)=[J,A+K_1SA^{\ast}SK_1-K_2SASK_2-K_1K_2A^{\ast}K_2K_1](x,t).
\end{gather}
Like in previous considerations we decompose the matrix $A(x,t)$
using two matrix factors $X(x,t)$ and $F(x,t)$ and derive some
dif\/ferential equation for $F(x,t)$ which leads to
\[F(x,t)=e^{i\lambda^{+}(Jx+It)}F_0.\]
The linear system for $X(x,t)$ in this case is following
\begin{gather}
\left(\openone-\frac{K_2SA(x,t)SK_2}{2\lambda^{+}}+\frac{K_1SA^{\ast}(x,t)SK_1}
{2i\nu}-\frac{K_1K_2A^{\ast}(x,t)K_2K_1}{2\mu}\right)SF(x,t)\nonumber\\
\qquad{}=X(x,t)\alpha(x,t),\label{BaseEq}
\end{gather}
where $\alpha(x,t)$ is a linear function of $x$ and $t$ as follows
\[
\alpha(x,t)=iF^T_0(Jx+It)SF_0.
\]
Starting from the equation (\ref{BaseEq}) by multiplying with
$K_{1}S$, $K_{2}S$ and performing complex conjugation if
necessary, we can derive the following auxiliary linear system
\begin{gather*}
SF(x,t)=X\alpha+Y\frac{(G,F)}{2\lambda^{+}}-Z\frac{(H,F)}{2i\nu}+
W\frac{(N,F)}{2\mu},\\
SG(x,t)=X\frac{(F,G)}{2\lambda^{+}}+Y\alpha+Z\frac{(H,G)}{2\mu}-
W\frac{(N,G)}{2i\nu},\\
SH(x,t)=X\frac{(F,H)}{2i\nu}+Y\frac{(G,H)}{2\mu}+
Z\alpha^{\ast}+W\frac{(N,H)}{2(\lambda^{+})^{\ast}},\\
SN(x,t)=X\frac{(F,N)}{2\mu}+Y\frac{(G,N)}{2i\nu}+
Z\frac{(H,N)}{2(\lambda^{+})^{\ast}}+W\alpha^{\ast},
\end{gather*}
 where we
introduced auxiliary entities
\begin{gather*}
Y(x,t)=K_2SX(x,t),\qquad Z(x,t)=K_1SX^{\ast}(x,t),\qquad W(x,t)=K_1K_2X^{\ast}(x,t),\\
G(x,t)=K_2SF(x,t),\qquad H(x,t)=K_1SF^{\ast}(x,t),\qquad
N(x,t)=K_1K_2F^{\ast}(x,t),\\ (F,H)=F^TSH.
\end{gather*}
 In matrix notations
this system reads
\[\left(\begin{array}{cccc}
SF,& SG, & SH, & SN
\end{array}\right)=\left(\begin{array}{cccc}
X,& Y, & Z, & W
\end{array}\right)\left(\begin{array}{cccc}
\alpha & a      &   b          &  c       \\
a      & \alpha &   c          &  b       \\
b^{\ast}   & c^{\ast}    & \alpha^{\ast}& a^{\ast} \\
c^{\ast}    & b^{\ast}   &  a^{\ast}    & \alpha^{\ast}
\end{array}\right),
\]
where
\[a(x,t)=\frac{(F(x,t),G(x,t))}{2\lambda^{+}},\qquad b(x,t)=\frac{(F(x,t),H(x,t))}{2i\nu},\qquad
c(x,t)=\frac{(F(x,t),N(x,t))}{2\mu}.\] To calculate $X(x,t)$ we
just have to f\/ind the inverse matrix of the block matrix shown
above. In the simplest case when
$\mbox{rank}X(x,t)=\mbox{rank}F(x,t)=1$ and $\alpha\equiv 0$ we
have
\[\left(\begin{array}{c} X\\ Y\\ Z\\ W\end{array}\right)=
\frac{1}{\Delta(x,t)}\left(\begin{array}{cccc}
0 & a^{\ast} & b & -c\\
a^{\ast} & 0 & -c & b\\
-b & -c & 0 & a\\
-c & -b & a & 0\end{array}\right)\left(\begin{array}{c} SF\\
SG\\SH\\SN\end{array}\right),
\] where
\[\Delta(x,t)=|a(x,t)|^2+b^2(x,t)-c^2.\]

Finally putting the result for $X(x,t)$ in (\ref{breather}) we
obtain the quadruplet solution
\begin{gather*}
  Q(x,t)=\frac{1}{\Delta}\left[
\left(a^{\ast}(x,t)K_2F(x,t)+b(x,t)K_1F^{\ast}(x,t)-
c(x,t)K_1K_2SF^{\ast}(x,t)\right)F^T(x,t)\right.\\
{} -K_2S\left(a^{\ast}(x,t)K_2F(x,t)+b(x,t)K_1F^{\ast}(x,t)-
c(x,t)K_1K_2SF^{\ast}(x,t)\right)F^T(x,t)SK_2\\
{} +K_1S\left(a(x,t)K_2F^{\ast}(x,t)-b(x,t)K_1F(x,t)-
c(x,t)K_1K_2SF(x,t)\right)F^{\dag}(x,t)SK_1\\
\left. {}-K_1K_2\left(a(x,t)K_2F^{\ast}(x,t)-b(x,t)K_1F(x,t)-
c(x,t)K_1K_2SF(x,t)\right)F^{\dag}(x,t)K_2K_1\right].
\end{gather*}

Consider the 4-wave system associated with the $\mathbf{B}_2$
algebra. Let $K_1=K_2=\openone$ then its generic quadruplet (or
breather-like) solution is
\begin{gather*}
q_{10}(x,t)  =  \frac{4}{\Delta}\mbox{Im}\left[a^{\ast}N_1
\cosh(\varphi_1+\varphi_2)-\frac{imN_1^{\ast}}{\mu\nu}\left(\mu\cosh(\varphi^{\ast}_1+\varphi_2)
-i\nu\cosh(\varphi^{\ast}_1-\varphi_2)\right)\right]N_2,\\
q_{11}(x,t)  =
\frac{4}{\Delta}\mbox{Im}\left[a^{\ast}N_1\sinh(\varphi_1)
-\frac{im\lambda^{+}}{\mu\nu}N^{\ast}_1\sinh(\varphi^{\ast}_1)\right]F_{0,3},\\
q_{12}(x,t)  =
\frac{4}{\Delta}\mbox{Im}\left[a^{\ast}N_1\cosh(\varphi_1-\varphi_2)
-\frac{imN^{\ast}_1}{\mu\nu}\left(\mu\cosh(\varphi^{\ast}_1-\varphi_2)
-i\nu\cosh(\varphi^{\ast}_1+\varphi_2)\right)\right]N_2,\\
q_{01}(x,t)  =
\frac{4}{\Delta}\mbox{Im}\left[a^{\ast}N_2\cosh(\varphi_2)
-\frac{im{\lambda^{+}}^{\ast}}{\mu\nu}N^{\ast}_2\cosh(\varphi^{\ast}_2)\right]F_{0,3},
\end{gather*}
where
\begin{gather*}
a(x,t)=\frac{1}{\mu+i\nu}\left[N^2_1\cosh 2\varphi_1+N^2_2\cosh
2\varphi_2 +\frac{F^2_{0,3}}{2}\right],\\
b(x,t)=\frac{m(x,t)}{i\nu},\qquad
c(x,t)=\frac{m(x,t)}{\mu},\\
m(x,t)=|N_1|^2\cosh (2\mbox{Re}\ \varphi_1)+|N_2|^2\cosh
(2\mbox{Re}\ \varphi_2)+\frac{|F_{0,3}|^2}{2},\qquad
N_{\sigma}=\sqrt{F_{0,\sigma}F_{0,6-\sigma}},\\
\varphi_{\sigma}(x,t)=i\lambda^{+}(J_{\sigma}x+
I_{\sigma}t)+\frac{1}{2}\log\frac{F_{0,\sigma}}{F_{0,6-\sigma}},
\qquad\sigma=1,2.
\end{gather*}

\section{Conclusion}\label{sec:4}

New $N$-wave type equations related to the orthogonal algebras and
obtained from the generic ones via additional reductions are
analyzed. These new nonlinear equations are solvable by the
inverse scattering method. In particular, the systems related to
the $\mathfrak{so}(5)$ algebra involve $4$ waves. Imposing
$\mathbb{Z}_2$ and  $\mathbb{Z}_2\times\mathbb{Z}_2$ reductions on
them we obtain $4$-wave systems of physical importance. The
$4$-wave system with a canonical $\mathbb{Z}_2$ reduction
describes Stokes-anti-Stokes Raman scattering. The
$\mathbb{Z}_2\times\mathbb{Z}_2$-reduced $N$ wave system possesses
real valued solutions while other types of $N$ wave equations have
complex valued solutions. This determines its mathematical
importance.

The soliton solutions of these integrable systems are parametrized
by two types of parameters: by the discrete eigenvalues of $L$ and
by the ``polarization'' vector $F_0$. Therefore the problem of
classifying all types of one-soliton solutions of the $N$-wave
equations is equivalent to that of classifying all possible types
of discrete eigenvalues and polarization vectors. When both
reductions are applied we have two dif\/ferent conf\/igurations of
eigenvalues for the Lax operator~$L$: doublets and quadruplets.
This situation is analogous to one encountered in the sine-Gordon
case, which was the reason to call quadruplet solitons
`breather'-like solutions \cite{Varna}.

\subsection*{Acknowledgements}
This work is partially supported by a contract 1410 with the
National Science Foundation of Bulgaria. This work has been
supported also by the programme ``Nonlinear Phenomena in Physics
and Biophysics'', contract 1879. We also thank the referees for
the careful reading of our paper.

\pdfbookmark[1]{References}{ref}
\LastPageEnding

\end{document}